\documentclass[a4paper,fleqn,usenatbib]{jec}

\usepackage{ae,aecompl}

\usepackage[latin9]{inputenc}

\usepackage{graphicx}
\usepackage{amsmath}
\usepackage{amsfonts}
\usepackage{amssymb}
\usepackage{color,hyperref}

\hypersetup{colorlinks,breaklinks,
           linkcolor=blue,urlcolor=blue,
           anchorcolor=blue,citecolor=blue}

\usepackage{algorithm}
\usepackage[noend]{algpseudocode}
\algrenewcommand\algorithmicindent{0.5em}%

\newcommand{\tab}[1]{Tab.~#1}
\newcommand{\fig}[1]{Fig.~#1}
\newcommand{\eq}[1]{Eq.~#1}

\usepackage{etoolbox}
\makeatletter
\patchcmd\@combinedblfloats{\box\@outputbox}{\unvbox\@outputbox}{}{\errmessage{\noexpand patch failed}}
\makeatother

\title[Adversarial training applied to CNN]{Adversarial training applied to  Convolutional Neural Network for photometric redshift predictions}

\author[J.E Campagne]{Jean-Eric Campagne$^{1}$\thanks{Email: campagne@lal.in2p3.fr}
	\\
$^{1}$Universit\'e Paris-Saclay, CNRS/IN2P3, IJCLab, 91405 Orsay, France\\
}
\pubyear{2019}

\begin{document} 
\label{firstpage}
\pagerange{\pageref{firstpage}--\pageref{lastpage}}
\maketitle

\begin{abstract}
The use of Convolutional Neural Networks (CNN) to estimate the galaxy photometric redshift probability distribution by analysing the images in different wavelength bands has been developed in the recent years thanks to the rapid development of the Machine Learning (ML) ecosystem. Authors have set-up CNN architectures and studied their performances and some sources of systematics using standard methods of training and testing to ensure the generalisation power of their models. So far so good, but one piece was missing : does the model generalisation power is well measured? The present article shows clearly that very small image perturbations can fool the model completely and opens the  Pandora's box of \textit{adversarial} attack. Among the different techniques and scenarios, we have chosen to use the Fast Sign Gradient one-step Method and its Projected Gradient Descent iterative extension as adversarial generator tool kit. However, as unlikely as it may seem these adversarial samples which fool not only a single model, reveal a weakness both of the model and the classical training. A revisited algorithm is shown and applied by injecting a fraction of adversarial samples during the training phase. Numerical experiments have been conducted using a specific CNN model for illustration although our study could be applied to other models - not only CNN ones - and in other contexts - not only redshift measurements - as it deals with the complexity of the boundary decision surface.  
\end{abstract}

%
%


 

\begin{keywords}
galaxies: distances and redshifts - surveys - techniques:photometric redshifts - methods: data analysis - statistical
\end{keywords}

%
\section{Introduction}
The outcomes of forthcoming very large optical surveys as LSST \citep{2008arXiv0805.2366I,Ivezi__2019} or red-optical and infrared surveys as Euclid \citep{2011arXiv1110.3193L}  will depend crucially up on the  galaxy  \footnote{The galaxies are not the only objects that will be catalogued and the conclusion to our study can nicely be generalized.} positions on the sky and especially along the  light of sight, i.e. the redshift. As the accurate spectroscopic redshift determinations are costly and time consuming such that they are limited to sub-samples of galaxies as in  BOSS (SDSS-III) \citep{2015ApJS..219...12A} and eBOSS (SDSS-IV) \citep{2016AJ....151...44D}  and the future DESI \citep{2013arXiv1308.0847L} or Euclid (spectro sub-sample) surveys, the majority of the galaxy redshifts of past surveys as DES Y1 \citep{2019MNRAS.489.5453S} and future as LSST and Euclid rely on the photometric measurements. The reconstruction performances of photometric redshift (hereafter named photo-z or $z_{phot}$)  has been extensively studied in the literature since the work of \citep{1962IAUS...15..390B} with different kinds of methods such as \textit{template-fitting}, \textit{feature based machine learning} and \textit{image based machine learning}.

%

The \textit{template-fitting} methods, first developped by \citep{1986ApJ...303..154L}, have been used for instance in the references \citep{1999MNRAS.310..540A,2000ApJ...536..571B,2006MNRAS.372..565F,2008ApJ...686.1503B} and recently in \citep{2014A&A...561A.128G, 2019A&A...623A..76A}. For a given galaxy, the algorithm matches the magnitude distributions in the different filter wide bands (e.g. \textit{ugriz}) to a suite of interpolate spectral energy distributions from few known galaxies and it results a photo-z value and error in a fully probabilistic fashion with explicit priors. 

The \textit{feature based machine learning} methods use different types of tools as Decision Trees (DT; \cite{Quinlan:1986:IDT:637962.637969}), Random Forests (RF; \cite{Breiman:2001:RF:570181.570182}), Support Vector Machine (SVM; \cite{Boser:1992:TAO:130385.130401}),  k-nearest neighbors (KNN; \cite{doi:10.1080/00031305.1992.10475879})  as well as Muti-Perceptron Layers (nicknamed either MLP or ANN; \cite{Werbos:74,Rumelhart1986LearningRB}). They have been used for instance in the references \citep{2004PASP..116..345C, 2005PASP..117...79W,2013MNRAS.432.1483C,2016PASP..128j4502S,10.1093/mnras/stw1009,2019A&A...623A..76A}. These methods use as input for a given galaxy the different magnitudes (or colors) measured in the different filters augmented eventually by other user driven information, and give as output a single photo-z value or a probability density distribution (p.d.f). But contrary to the \textit{template-fitting} methods, the prediction is based on the determination of the internal parameters of each tool during a supervised preliminary phase, where a sub-sample of galaxies with known redshifts are processed. The later redshifts considered as the \textit{true} values are mostly known thanks to the accurate spectroscopic measurements. The \textit{feature based machine learning} methods have shown better prediction accuracy than the  \textit{template-fitting} methods although a combination of the two class of methods has been investigated for instance in reference \citep{2017MNRAS.466.2039C}. The \textit{feature based machine learning} methods rely explicitly on manual feature extractions that may not capture all the information present in the images: for example the point spread function (PSF) variations are problematic \citep{2012MNRAS.421.2355H,2015MNRAS.453.1136J, 2019A&A...621A..26P}.

The last type of methods, i.e. \textit{image based machine learning},  have been investigated to overcome these drawbacks, and their popularity has grown in the recent years thanks to the Machine Learning ecosystem developments. Notably, the success of the deep convolutional neural networks (CNN) since the pioneer work at the end of the 80's and the 90's \citep{NIPS1989_293,44e2afaa580a48bc8b13633b22ff10b4,cccfa4f7238441b0a9021bb9f917e8ed} has dramatically changed  the paradigm in many fields among the  computer vision (image and video classification, segmentation), the speech recognition, the natural language treatment \citep{LecBen15}. In brief, a CNN is a feed-forward neural network with two main parts: the first one is composed of several blocks of convolutional layers followed by sub-sampling layers, while the second one is composed of fully connected (or dense) layers from which the last layer gives the desired output (e.g. redshift p.d.f). As it is common to say, during the training phase, the first part somewhat learns the optimal feature representation which replaces the manual features extraction of the  previous methods, while the second part is the classifier. For   photometric redshift predictions, one can read for intance the following references \citep{2018A&A...609A.111D, 2019A&A...621A..26P}. Notice that in astronomy CNNs have been also used for other tasks as  image processing task (e.g. \cite{2016arXiv161204526F}) among which the deblending (e.g. \cite{2019MNRAS.485.2617R,2019MNRAS.tmp.2430B}), and also object classifications (e.g. \cite{Kim:2016knv,refId0,2018A&C....25..103G}).

The  \textit{image based machine learning} methods for photometric redshift predictions do suffer from systematics among which some are shared with other supervised methods as the representativeness quality and size of the training samples and the quality of the input information. The later source of systematics are closely related to the images themselves which may influence the prediction:  for instance the presence or not of neighbouring objects even if there do not blend the central object, the possible orientation of the object, its possible non central position, the non uniformity of the PSF, the different filter response calibrations, and so on. All these source of systematics are of course relevant, but this article is motivated by the existence of intrinsic perturbations of the images that can fool the classifier prediction, i.e the photometric redshift p.d.f as discussed in the present article. The subject known as \textit{adversarial examples} in the ML literature has been first studied by \citep{Szegedy2013IntriguingPO, 2014arXiv1412.6572G}, it was one of the challenges proposed during the NIPS '17 Competition \citep{10.1007/978-3-319-94042-7_11,2018arXiv180400097K} and it gave rise to communications in later NISP conferences as it is an important subject on ML robustness and security\footnote{See \url{https://nicholas.carlini.com/writing/2019/all-adversarial-example-papers.html} to get an idea of the literature growth about this subject.}. 

Contrary to the expected good generalisation power of trained  machine learning models with good practices and with good representativeness of the different training/testing/validation sets, the  \textit{adversarial examples} show dramatic prediction failures while the perturbations of the inputs are so tiny that one  expects very marginal impacts. Unfortunately, this is not the case and it has been realised that one should envisage countermeasures.  The different scenarios studied in the ML literature are:  \textit{target} or \textit{non-target} attack which distinguishes between image perturbations that fool the model predictions either by forcing the model towards a predefined redshift value either just by forcing the model to not-predict the correct redshift; and depending on the knowledge of the architecture of the model, one qualifies the attack as \textit{white box}-like (full knowledge) or \textit{black box}-like (no real knowledge of the model). Put in the context of photometric redshift predictions,  we think that \textit{target} attacks are not to be envisaged in our case as if there exist input perturbations  that fool the model, we do not think that they are so malicious. Concerning white versus black box scenarios, we think that white box ones are valuable as they give us some chance to propose a new training method that leads to better immunity against image perturbations. However, it will be shown a black box like effect using adversarial examples of a model given as inputs to a different model architecture.

The present article is organised as followed. Section~\ref{sec:data-cnn} summarises the data and CNN models used. Essentially, we are using the \textit{Inception} CNN proposed in reference \citep{2019A&A...621A..26P} which has shown good redshift prediction performances and serves as reference for our studies. We have also chosen this CNN model because the present article is rather a proof of concept which needs a CNN of known architecture details with good performances, well trained in the classic sense and the possibility to use the same set of input data. Section~\ref{sec:adv-gene} introduces the adversarial image generation and shows their dramatic effects, and explains what are the effects of these samples telling us about? Then, in Section~\ref{sec:adv-training} a new training algorithm is detailed as well as the performances of \textit{Inception} retrained model depending on the adversarial generator even if the outcomes are not perfect. Finally, in Section~\ref{sec:conclusion} we summarize our main results.
\section{Data and CNN model}
\label{sec:data-cnn}
\begin{figure}
\centering
\includegraphics[width=\columnwidth]{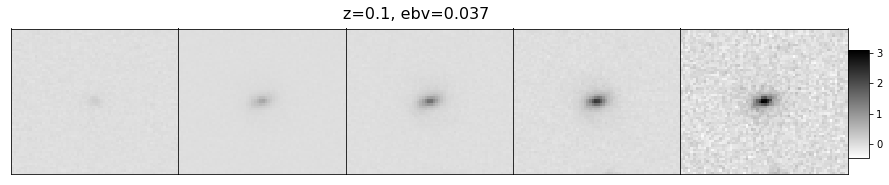}\\
\includegraphics[width=\columnwidth]{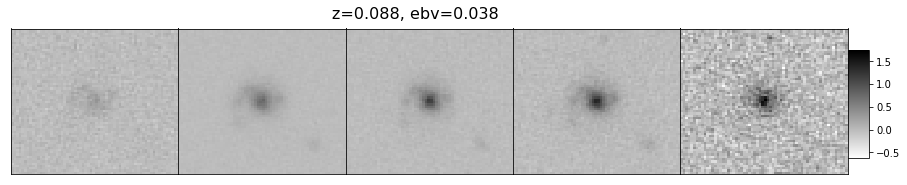}\\
\includegraphics[width=\columnwidth]{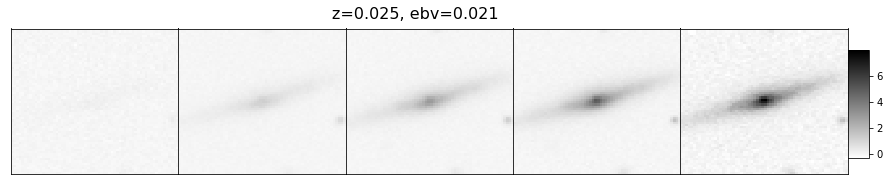}
\caption{Examples of the $5\times 64\times 64$ pixel cube as input of the CNN. For each imaged galaxy we show the \textit{ugriz} filter responses and on top the $z_{spec}$ value as well as the $E(B-V)$ redenning correction. Each series of filter frames has its own LUT shown on the right side.}
\label{fig-imgs-sdss}
\end{figure}

\begin{figure}
\centering
\includegraphics[width=0.75\columnwidth]{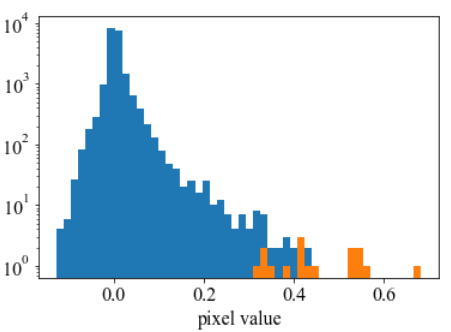}\\
\caption{Example of the distribution of the pixel values of the five filter patches of a single image with in orange the values of the central galaxy pixels located at $z=0.27$.}
\label{fig-pixel-val}
\end{figure}

The numerical experiments undertaken in this article rely mostly on the model and data used in \citep{2019A&A...621A..26P} and we only briefly introduce these key ingredients. The data are originated from the SDSS multi-band imaging and spectroscopic redshift survey using  a  dedicated  2.5-meter  telescope  at  Apache  Point Observatory in New Mexico. The authors used the data release 12 (DR12; \cite{2015ApJS..219...12A}) from which has been extracted the sources classified as galaxy with spectroscopic redshifts mostly in the $[0, 0.3]$ range. The  'corrected frames'  image of the SDSS \textit{ugriz} filters of data release 8  have been treated by the authors using the \verb|SWARP| code\footnote{\url{http://astromatic.net/software/swarp}} \citep{2002ASPC..281..228B} to resample to a common pixelgrid and stack all the available image data. It results of a $5\times 64\times 64$ pixel cube centred on the galaxy coordinates (\fig{\ref{fig-imgs-sdss}}). The pixel values are float values possibly negative, this is in contrast to usual RGB images used in general by ML designers (eg. CIFAR images for data challenges \footnote{https://www.cs.toronto.edu/~kriz/cifar.html}). An example is shown in \fig{\ref{fig-pixel-val}}. 
Besides the images, the $E(B-V)$ ('ebv' variable) redenning correction  in magnitudes at the position of each object is also used as extra variable given to the fully connected part of the network.

Concerning the CNN architecture, the authors have used an \textit{Inception} model inspired from GoogleLeNet \citep{2014arXiv1409.4842S}. Originally, GoogleLeNet has been used for the ImageNet Large-Scale Visual Recognition Challenge 2014 and the architecture was motivated by mainly the following arguments: first, to deal with size and location variations of the information in the input  image, one needs small kernel size for localized information and larger kernel for more distributed information; second, very deep networks are more prone to overfitting, so one needs to keep the network as short as possible;  third, stacking a lot of convolution layers tends to be computationally intensive, this also tends to keep the network as short as possible. These arguments remains true concerning the information variations, and also if the hardware and algorithmic codes have been improved since 2014 and allows very deep architecture as the Resnet series \citep{2015arXiv151203385H}, for this study keeping architecture simple is suited to feed a single GPU and keep the training phase in reasonable time-scale (i.e. not larger than 48 hours). But,  anticipating to the result of the Section~\ref{sec:adv-training}, one has to keep in mind that the classification power of the network also has to be taken into account to design the architecture.

So, the model of \cite{2019A&A...621A..26P} is composed of a $5\times 5$ kernel convolutional  layer followed by a series of 5 inception cells with $1\times 1$, $3\times 3$ and $5\times 5$ kernels, and finally a pair of fully connected layers to transform the learned features into a prediction. Instead of using a max pooling layer after a convolution due to possible large single pixel intensity variations, one prefers to use average pooling.   The activation functions for the fully connected layers are ReLU, while for the convolutional layer one applies a PReLU activation with a learnable parameter.  Some details upon the size of the different layers are given in Appendix \ref{app-inception}. In Section~\ref{sec:adv-training}, we examine some modifications of this model. The inputs are typically batches of 64 samples\footnote{It is implicit that a sample is composed of the collection of the \textit{ugriz} filter datacubes associated with the 'ebv' variables.} and the output is a vector of 180 bins to sample photometric redshift p.d.f prediction  taking into account that $z_{spec} \in [0, 1]$ (i.e. the bin width is $5.5\times 10^{-3}$ small enougth to sample the photometric redshift p.d.f\footnote{\cite{2019A&A...621A..26P} use a bin width of   $2.2\times 10^{-3}$ but the redshift p.d.f sampling is not affected by a larger bin width as it is shown on \fig{\ref{fig-advex-fgsm}}.}). 

For comparison purpose, a  simpler CNN has been set-up consisting of 1 convolutional $5\times 5$ kernel layer followed by 2 convolutional $3\times 3$ kernel layers and 2 fully connected layers. Details can be seen on Appendix~\ref{app-inception}. Notice that the 'ebv' variable is also used at the entrance of fully connected part of the network. This simpler CNN model gives satisfactory results compared to the \textit{Inception} model, but it was not intended to outperform the later.

As in the following sections we need to train the different models in different configurations, so we could not use the pre-trained \textit{Inception} model available on github\footnote{\url{https://github.com/jpasquet/photoz}}, and
we have used the latest \verb|pyTorch| library\footnote{\url{https://pytorch.org/}} v1.3.0  \citep{2019arXiv191010775O} from which we have changed the default weight initialisations into the Xavier's uniform method and the bias are initialised to 0.1, to follow initialisation conditions used by \cite{2019A&A...621A..26P} .

For the minimisation process we use the cross-entropy loss between the CNN output (i.e. the 180 bins vector) and the spectroscopic redshift value converted to a hot vector of same size\footnote{That is to say a vector of length 180, with all the components  set to 0, except the one corresponding to the $z_{spec}$ value which is set to 1.}. Concerning the optimiser, we have performed numerical studies with the traditional SGD \citep{Sutskever:2013:IIM:3042817.3043064} and  Adam codes \citep{2014arXiv1412.6980K}, as well as the recent Adam improvement code \citep{2017arXiv171105101L} called AdamW which is available in \verb|pyTorch| since version 1.2.0. Notice that both for Adam and AdamW, we have set the $\epsilon$-parameter to $0.1-1.0$ values much higher than the default value (i.e. $10^{-8}$), first because it is recommended for Inception architecture in the  \verb|TensorFlow|  library, and second by our own experience below $0.01$ the training manifests large instabilities. The initial learning rate is typically set to $0.01$, and we have used a reduce learning rate scheduler to decrease by 20\% the learning rate if after 5 or 10 epochs the test accuracy has not been improved.

From the initial $659\, 857$ input samples, after random shuffling,  we have set-up different sets of $100\, 000$ samples, one common for the test set and several for different training sets, and also a different set of $10\, 000$ samples for the adversarial test experiments. After been gathered in batches, we  randomly apply a combination of geometrical  transformations such as horizontal and vertical flips and $90\degr$, $180\degr$ and $270\degr$ rotations. The number of training epochs is typically 150, but first sometimes the minimization has required early stopping and relaunching with a different optimiser settings, and second we have pushed up to 300 epochs for some special training tests. All the numerical experiments have been conducted at the CC-IN2P3\footnote{\url{https://cc.in2p3.fr/en/}} GPU-Nvidia farm from which a single Tesla V100 is used at a time. 
\section{Adversarial images generation}
\label{sec:adv-gene}
\subsection{Classical training of \textit{Inception} model}
\begin{figure}
\centering
\includegraphics[width=0.8\columnwidth]{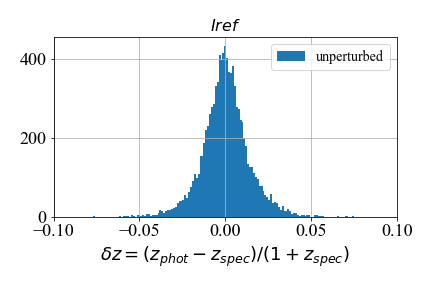}\\
\includegraphics[width=0.8\columnwidth]{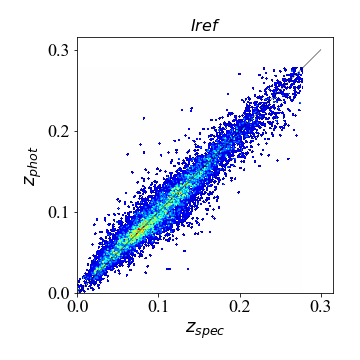}
\caption{Results obtained with the the reference \textit{Inception} model  trained and tested with unperturbed images: (top) distribution of the $\delta z$ variable; (bottom) $z_{phot}$ versus $z_{spec}$.}
\label{fig-dz-ref}
\end{figure}

First of all, an \textit{Inception} model, nickname hereafter \textit{Iref} has been trained "classically", that is to say using a pair of train/test sets and tuning the optimizer hyper-parameters to get the training and testing losses as close as possible up to an epoch where the network starts to overfit, that is to say when the test loss increases while the train loss still decreases. The results are presented in terms of the distribution of the variable $\delta z = (z_{phot}-z_{spec})/(1+z_{spec})$  (\fig{\ref{fig-dz-ref}} top) and the $z_{phot}$ versus $z_{spec}$ comparison (\fig{\ref{fig-dz-ref}} bottom). For the sake of comparison between different numerical experiments, we have used the following variables: 
\begin{itemize}
\renewcommand{\labelitemi}{\scriptsize$\bullet$}
\item the bias defined as the mean of the  $\delta z$ distribution;
\item the $\sigma_\mathrm{MAD} = 1.4826 \times |\delta z - \mathrm{Median}(\delta z)|$;
\item and the fraction $\eta$ of outliers such that $|\delta z| > 0.05$.
\end{itemize}
The results concerning \textit{Iref} are presented on \tab{\ref{tab-non-robust}} and for the "classical" training with non perturbed images (i.e. non adversarial images) the reader should have a look at the line 'non perturbed'. The obtained results are satisfactory enough and comparable with the ones obtained in \citep{2019A&A...621A..26P}, so it can serve as a rather good reference starting point.
\subsection{Adversarial samples}
Let us now get a brief introduction to adversarial samples generation. In a classical supervised training, the parameters $\theta$ of a model $f$ are adjusted to minimise an empirical risk, thanks to the set of $\{x_i,z_i\}_{i\leq N}\in D_{train}$ samples, where  $x_i$ is an input image\footnote{Notice that the 'ebv' variable is implicitly added but it remains unperturbed in the following numerical experiments.} and $z_i$ is the spectroscopic redshift considered as the true value. The expression of the empirical risk is
\begin{align}
R_e(f_\theta,D_{train}) &= \frac{1}{|D_{train}|} \sum_{(x,z)\sim D_{train}}\ell(f_\theta(x),z)  \nonumber \\
&= \frac{1}{N}\sum_{i=1}^N \ell(f_\theta(x_i),z_i). 
\label{eq-normal-train}
\end{align}
with $\ell$ the cross-entropy loss function. Depending on the optimizer (SGD, Adam, and so on), the algorithm is different but for the sake of definiteness and illustration, we use here a Batched Gradient Descent method  where $(x,z)$ samples are gathered on  batches named $B_{train}$. After an initialisation phase, the values of the $\theta$ parameters are updated step-by-step as: 
\begin{equation}
\theta_{t+1} = \theta_t - \alpha\ \frac{1}{|B_{train}|}\sum_{(x,z)\sim B_{train}} \left[ \nabla_\theta \ell(f_\theta(x),z)\right]_{\theta=\theta_t}
\label{eq-normal-gd}
\end{equation}
with $\alpha$ the \textit{learning rate}. 
Now, the adversarial samples are images to which is added a small amount, noted $\delta$ according to
\begin{multline}
R_{adv,e}(f_\theta, D_{train})=\\ \frac{1}{|D_{train}|}\sum_{(x,z)\sim D_{train}} \left[\underset{\|\delta \|\leq \varepsilon}{\mathrm{max}}\ \ell(f_\theta(x+\delta ),z)\right].
\label{eq-Radv-e}
\end{multline}
The minimisation of  $R_{adv,e}$ to get the $\theta$ parameters needs to solve a \textit{minimax problem} also known as a \textit{saddle point problem}, with an inner maximization part included in the more familiar outer minimisation part. 

The maximisation problem leads to several kinds of methods that generate adversarial examples. Such kind of generating method is called an \textit{attack} in the literature and we  occasionally use this terminology later on. The first method, called the Fast Sign Gradient Method (FSGM), has been proposed in \citep{2014arXiv1412.6572G} and more elaborated  attacks are detailed for instance in the following references  \citep{2016arXiv161101236K,2017arXiv171207107Y,2017arXiv170606083M,2018arXiv180400097K,10.1007/978-3-319-94042-7_11,2019arXiv190502175I}. In the following, we briefly summarize the FSGM attack which is simple and powerful, as well as one of its  generalisation, i.e.  the Projected Gradient Descent (PGD)\footnote{This method is also called Basic Iterative Method in the literature.} iterative method which is more aggressive.
\subsection{The FSGM one-step generator}
The  solution of the  maximisation problem (Eq.~\ref{eq-Radv-e})  is noted $\delta^\ast$, it generally depends on $x$ and satisfies:
\begin{equation}
\delta^\ast(x) = \underset{\| \delta\| \leq \varepsilon }{\mathrm{argmax}}\ \ell(f_\theta(x+\delta ),z).
\label{eq-deltaast-general}
\end{equation}
The maximisation ensures that perturbations smaller than $\delta^\ast$ in terms of the norm used, have smaller impact on the model predictions. In the following, the infinity norm (L$_\infty$) is used as it gives good results, but for the sake of completeness a numerical experiment has been conducted with the L$_2$ norm\footnote{As a remainder, consdering a vector $x=(x_1,\dots,x_n)\in\mathbb{R}^n$, $\|x\|_2 = (x_1^2\dots+x_n^2)^{1/2}$, while $\| x \|_\infty = \mathrm{sup}(|x_1|,\dots,|x_n|)$.} giving more sparse perturbations\footnote{Notice that models trained with adversarial samples generated by the infinity norm give better robustness \citep{2017arXiv170606083M}.}.  In case of the infinity norm $\| \delta \|_\infty\leq \varepsilon$ and under some hypothesis mentioned later, the solution reads:
\begin{equation}
\delta^\ast(x) = \varepsilon\times \mathrm{sign}\left( \nabla_\delta \ell(f_\theta(x+\delta),z) \right)
\label{eq-deltaast-linear}
\end{equation}
from which the FSGM name is originated, i.e. Fast Sign Gradient Method. 

\begin{table}
\caption{Results presented in Sec.~\ref{sec:adv-gene} considering a classical training of \textit{Inception} like models as one for reference \textit{Iref} and four twins  \textit{I0}-\textit{I3} (\tab{\ref{tab-inception-model}}), and a completely different CNN model (\tab{\ref{tab-cnn-model}}). These five alternate models are used to process adversarial samples of the \textit{Iref} model. The three  variables (bias, $\sigma_\mathrm{mad}$, and the fraction of outliers $\eta$) are computed from the $\delta_z = (z_{phot}-z_{spec})/(1+z_{spec})$ distributions. The true bias value should be multiplied by $10^{-4}$, similarly the  $\sigma_\mathrm{mad}$ should be multiplied by $10^{-3}$ and  $\eta$ is expressed in percentage. The results are presented for non perturbed images as well as all the investigated adversarial image generators as Fast Sign Gradient Method and Projected Gradient Descent, considering by default the infinity norm, except a the last part of the table where the L$_2$ norm is applied. In this later case, the $\varepsilon$ of \eq{\ref{eq-P2-projection}} is  $ \varepsilon^\ast\times (2n/e\pi)^{1/2}$ with $n=64^2$ to take into account the difference of  $\varepsilon$-ball volumes between the L$_2$ and infinity norm. Notice, for adversarial attack results, the bias values are not shown as they are not relevant since the gaussian-like distribution of $\delta_z$ is no more preserved.}
\label{tab-non-robust}
\centering
\begin{tabular}{lccc} \hline \hline

Models (images) & bias ($\times 10^{-4}$) & $\sigma_\mathrm{mad}$ ($\times 10^{-3}$) & $\eta$ (\%) \\ \hline
 \textit{Iref} (non perturbed)  & $0.3$ & $11$ & $1$ \\ \hline 
 \multicolumn{4}{c}{ $\varepsilon=10^{-2}$, Single FSGM} \\
  \textit{Iref}  & -- & $66$ & $42$ \\
  \textit{I0}-\textit{I3} (\textit{Iref} adv) & -- & $[63,68]$ & $[40,43]$ \\
 CNN (\textit{Iref} adv) & -- & $76$ & $49$	\\ \hline
  \multicolumn{4}{c}{$\varepsilon=10^{-2}$, PGD,  $\alpha=10^{-3}$, $n_{iter}=10$} \\
  \textit{Iref}  & -- & $82$ & $59$ \\
 \textit{I0}-\textit{I3} (\textit{Iref} adv) & -- & $[76,78]$ & $[52,55]$ \\   \hline



 \multicolumn{4}{c}{$\varepsilon^\ast=10^{-2}$, PGD L$_2$-norm,  $\alpha=10^{-3}$, $n_{iter}=40$} \\
  \textit{Iref}  & -- & $79$ & $57$ \\
 \textit{I0}-\textit{I3} (\textit{Iref} adv) & -- & $[77,79]$ & $[53,55]$ \\   \hline
\end{tabular}
\end{table}

\begin{figure}
\centering
\includegraphics[width=0.8\columnwidth]{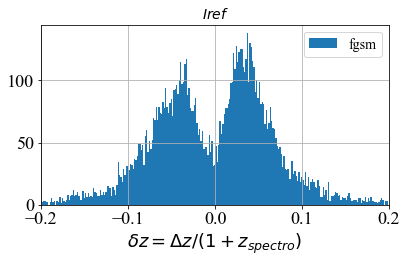}\\
\includegraphics[width=0.7\columnwidth]{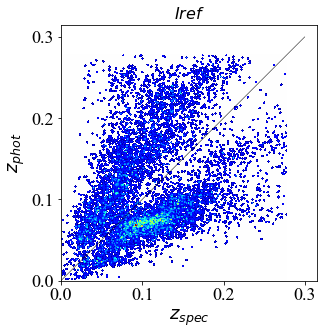}
\caption{Same kind of histograms (same binning) of \fig{\ref{fig-dz-ref}} but after a FSGM attack with $\varepsilon=0.01$.}
\label{fig-dz-fgsm}
\end{figure}

\begin{figure}
\centering
\includegraphics[width=\columnwidth]{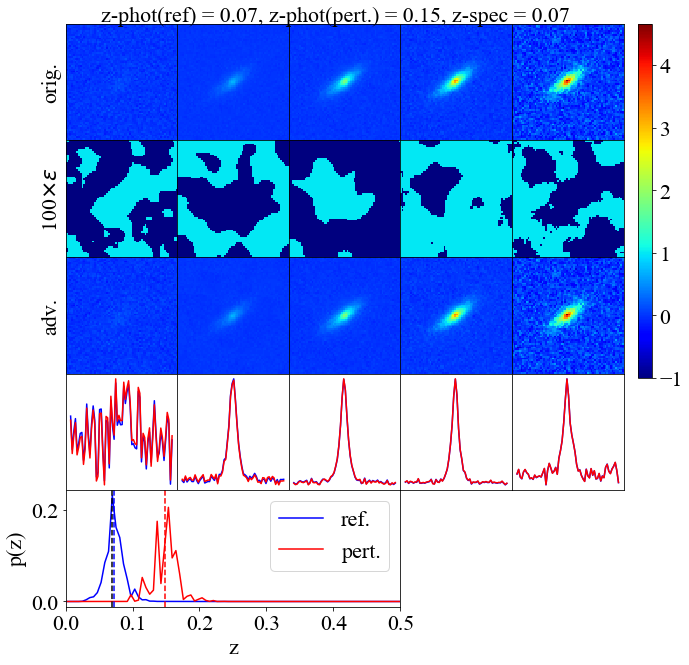}
\caption{Example of the impact of a FGSM attack ($\varepsilon = 10^{-2}$). From top to bottom: on the 1st line are displayed the 5 filter frames of an non perturbed original/reference image ($x$); on the 2nd line the corresponding frames of the perturbation $\delta^\ast(x)$ magnified by a factor 100 to be visible; on the 3rd line for the sake of completeness are shown the results $x+\delta^\ast(x)$; on the 4th line are shown the horizontal slices at the vertical middles of the 5 frames from $x$ in blue and $x+\delta^\ast(x)$ in red, notice in the left most panel that the perturbation as expected is far smaller than the background level; on the last line are shown in blue and red curves the predicted redshift p.d.f  for the non perturbed and the perturbed image as well as the blue and red vertical dashed lines corresponding to the $z_{phot}$ estimations as weighted means of the bin centre locations, and the black vertical dashed line corresponding to the $z_{spec}$ "true" redshift value. The color scale is common to all the filter frames.}
\label{fig-advex-fgsm}
\end{figure}

To see the impact of such perturbation added to image $x$, we have first conducted a numerical study to determined $\varepsilon$. When adding a random uniform noise to an original image $x$, i.e. $\delta(x) \sim \mathcal{U}[-\epsilon, +\epsilon]$, if $\varepsilon = 0.01$ there is absolutely no effect on the generalisation power of \textit{Iref} (i.e. the bias, $\sigma_{MAD}$ and $\eta$ variables remain unchanged), while if $\varepsilon = 0.1$  then the $\sigma_{MAD}$ is increased by 50\% and the number of outliers is multiplied by 3. So, we have fixed $\varepsilon = 0.01$ to perform an FSGM attack on \textit{Iref}. The results of such attack are displayed both in \fig{\ref{fig-dz-fgsm}} that can be compared to \fig{\ref{fig-dz-ref}} and in \tab{\ref{tab-non-robust}} (line "$\varepsilon=10^{-2}$, Single FSGM").
On a single image, \fig{\ref{fig-advex-fgsm}} shows the kind of perturbation generated by such attack.

As one clearly can appreciate, the very small perturbation generated by the attack has dramatic impacts on the photo-z predictions. But more than that, in \tab{\ref{tab-non-robust}}, one can see that the adversarial sample of the reference \textit{Inception} model impact not only other \textit{Inception} models trained with different seeds as well as different training sets, noted \textit{I0}-\textit{I3}, but also the different CNN model. Some remarks can be drawn from this  attack experiment: first, adversarial samples of one model also affect other  models which implies by consequence that one cannot use several models to combine their  results to cancel this underlying systematic error; second, the adversarial samples of one type of model architecture  also affect other model architectures. The later remark has been  identified in reference \citep{Szegedy2013IntriguingPO} and allows one for some applications to put a corner in the defence of a ML model (i.e. \textit{black box} attack). So, all this tends to suggest that adversarial samples are quite special  samples as they clearly break the traditional estimation of the generalisation power of a model using test and validation samples.  

The first reaction which was probably a common sense when these adversarial samples were discovered in the ML world, relied on the fact that the FSGM generated perturbations are so special that the probability of occurrence should be close to 0. But, this is not so simple even if we exclude malicious intent in this work. First, the level of the FSGM perturbation is so tiny that one cannot totally exclude that such perturbation could not be made during the processing of the different codes involved in a image treatment pipeline (i.e.  CCD pixel correction, deflating, de-biasing, calibration, stacking, resampling, cosmic tracks removal and so on\footnote{See for instance the different steps of the pipeline detailled in reference \cite{2018PASJ...70S...5B}.})  even if these codes are very sophisticated, especially if the experimental set-up  evolves in time or if the observation conditions are not 100\% of perfect photometric nights. Although, one can argue that these artefacts can be well mastered thanks to the long experience accumulate in image treatments. Second, intrinsically these adversarial samples reveal a weakness of the training and also probably of the model design. However,  in the Section \ref{sec:adv-training} we show how one can take advantage of the adversarial generation to perform a new training.  
\subsection{The PGD multi-steps generator}
Before exploring the adversarial samples benefit, let us explore  an other attack method which generalises in a simple manner the FSGM one. In fact, the $\delta^\ast$ expression \eq{\ref{eq-deltaast-linear}} is only exact in the linear case where $f_{\theta=(w,b)}(x) = w^T x + b$ \citep{2014arXiv1412.6572G}, even if this solution is nicely very effective as we have experienced. In the non-linear general case, one needs to use an algorithm to get an approximated solution of \eq{\ref{eq-deltaast-general}}. To do so, one can use a hill climbing algorithm. After an initialisation phase $\delta=\delta_0$, one performs a certain number of iterations ($n_{iter}$) to update $\delta$ as followed:
\begin{equation}
\delta \leftarrow \delta + \underset{\|u\|\leq \alpha}{\mathrm{argmax}} \left[ u^T. \nabla_\delta \ell(f_\theta(x+\delta),z)\right]
\end{equation}
which in case of the infinity norm leads to
\begin{equation}
\delta \leftarrow \mathcal{P}^\infty_\varepsilon\left[\delta + \alpha\ \mathrm{sign}(\nabla_\delta \ell(f_\theta(x+\delta),z))\right]
\label{eq-delta-iter-linfty}
\end{equation}
with $\alpha$ a \textit{learning rate} different from the one used for the $\theta$ minimisation part and with $ \mathcal{P}^\infty_\varepsilon$ the projection on the $\varepsilon$-L$_\infty$ ball such that $\| \delta \|_\infty \leq \varepsilon$. As one recognises, in some sense, the method applies iteratively an FSGM attack with a small step size  $\alpha$ decoupled from the $\varepsilon$ value. In the literature, this method is part of a series known as  Projected Descend Gradient attacks (PGD). Notice that a refinement of the above method consists of randomly initialise $\delta_0$ in the   $\varepsilon$-L$_\infty$ ball, however it has shown very little impact in our numerical experimental cases. There exists other methods to solve \eq{\ref{eq-deltaast-general}}, one is using a neural network which learns how to generate adversarial samples \citep{Jang_2019_ICCV}.

\begin{figure}
\centering
\includegraphics[width=0.8\columnwidth]{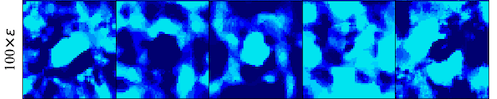}
\caption{Pattern of a PGD adversarial iterative attack with $\varepsilon=10^{-2}$ operating on the same original image of \fig{\ref{fig-advex-fgsm}} (with the same LUT). By comparison, one can see that such iterative method leads to smoother pattern.}
\label{fig-advex-pgd-linf}
\end{figure}

Considering $n_{iter} = 10$, \tab{\ref{tab-non-robust}} presents the results of the PGD attacks conditioned by $\alpha=\varepsilon/10$ both on the reference \textit{Inception} model as well as its four twins \textit{I0-I3}. See \fig{\ref{fig-advex-pgd-linf}} to have an idea of the perturbation pattern (i.e. $\delta$) generated by the PGD generator method. 
As anticipated, greater is the number of iterations, greater is the negative impact on the photo-z prediction. Although the effect is not proportional to the number of iterations as using $n_{iter}=20$ yields to an increase by 10\% of the $\sigma_{MAD}$ value and the number of outliers. One notices that the FSGM method, in a single step, already points towards the weakness of the network and the training.

Before diving into the robust training algorithm, for the sake of completeness, let us investigate what happens if we use the L$_2$ norm instead of the infinity norm. The $\delta^\ast$ solution \eq{\ref{eq-delta-iter-linfty}} is replaced by:
\begin{equation}
\delta \leftarrow \mathcal{P}^2_\varepsilon\left[\delta + \alpha\ \frac{\nabla_\delta \ell(f_\theta(x+\delta),z)}{\| \ell(f_\theta(x+\delta),z)\|_2}
\right]
\label{eq-delta-iter-l2}
\end{equation}
where the projection $\mathcal{P}^2_\varepsilon$ on the $\varepsilon$-L$_2$ ball is defined according to
\begin{equation}
 \mathcal{P}^2_\varepsilon(x) = \varepsilon\ \frac{x}{\mathrm{max}(\varepsilon, \|x\|_2)}.
 \label{eq-P2-projection}
\end{equation}
The results on the reference \textit{Inception} and its four twins are written on the last lines of \tab{\ref{tab-non-robust}}. Notice that to conduct such numerical experiment with L$_2$ norm in a comparable manner to the infinity norm, one needs to scale $\varepsilon$ according to the change of $\varepsilon$-ball volume between the two norms, and to get an similar impact on the photo-z prediction, we were forced to push the number of iterations up to $n_{iter}=40$. At this price,  L$_2$ norm can also clearly lead to negative impact on the photo-z predictions and the pattern of the perturbation is little sparser than with the infinity norm. But, this norm is significantly more time consuming with a not so different behaviour compared to the FSGM or the PGD methods. So, it will not be used hereafter in the rest of this article.

\section{Adversarial training}
\label{sec:adv-training}
\subsection{The algorithm}
In the previous section, we have shown that a simple adversarial sample generator, i.e. the FSGM attack with very small $\varepsilon$ value far below the image background, can lead to dramatic impact on the photo-z predictions, even if the \textit{Inception} model has been correctly trained, in a classical sense. At this stage, one may just forget this adversarial samples advocating a very unlucky probability to occur. But as we have argued, one may have a more proactive attitude taking as fact that the model training as a certain amount of  weakness due to our inability to understand the exact essence of the features that CNNs are really capturing \citep{2019arXiv190502175I}. Even if in photo-z prediction domain, the security is not engaged as it can be the case in cryptography or autonomous vehicle for instance, we envisage the second attitude, especially because we can strengthen the training. 
\begin{algorithm}
\caption{Adversarial training}
\label{algo-1}
\begin{algorithmic}[1]
\State Choose adversarial samples fraction and the attack generator (FSGM/PGD, $\varepsilon$, $\alpha$, number of iterations)
\State Do $\theta$ (model weights) initialisation 
\ForAll{mini-batch $B_{train}$}
\State $g \leftarrow 0$  \Comment{loss gradient w.r.t $\theta$}
\ForAll{$(x,z)\in B_{train}$}
\If{$x$ counts for an adversarial sample}, according to initial generator choice, find $\delta^\ast$:
\State $\delta^\ast \leftarrow \varepsilon\ \mathrm{sign}\left( \nabla_\delta \ell(f_\theta(x+\delta),z) \right)]$ \Comment{(FSGM)}
\State $\delta \leftarrow \mathcal{P}^\infty_\varepsilon\left[\delta + \alpha\ \mathrm{sign}(\nabla_\delta \ell(f_\theta(x+\delta),z))\right]$ \Comment{(PGD)}
\Else
\State $\delta^\ast \leftarrow 0$
\EndIf
\State$g \leftarrow g + \nabla_\theta \ell(f_\theta(x+\delta^\ast),z)$ \Comment{Update loss gradient}
\EndFor
\State $\theta \leftarrow \theta - \alpha \frac{g}{|B_{train}|}  $ \Comment{Update model weights}
\EndFor
\end{algorithmic}
\end{algorithm}

Then, let us elaborate a new training algorithm. In fact to minimize the empirical adversarial risk (\eq{\ref{eq-Radv-e}}), one can use a gradient descent method similar to \eq{\ref{eq-normal-gd}} but adapted to the adversarial empirical risk. This would lead to the following natural modified version:
\begin{align}
\theta_{t+1} &= \theta_t \nonumber \\
&- \alpha\ \frac{1}{|B_{train}|}\sum_{(x,z)\sim B_{train}} \nabla_\theta\left[\underset{\| \delta\|\leq \varepsilon}{\mathrm{max}}\ \ell(f_\theta(x+\delta ),z)\right]_{\theta=\theta_t}.
\end{align}
However, there are two remarks that can be raised:  the first one concerns the method to compute the gradient regarding the maximisation inner problem, the other  one pinpoints the fact that only adversarial samples are considered for the loss computation. Concerning the gradient computation, the difficulty is bypassed as followed: in a convex problem under some mathematical hypothesis, one can use the J. Danksin's theorem \citep{danskin1966theory} which yields  the simple result:
\begin{equation}
\nabla_\theta\left[\underset{\| \delta\|\leq \varepsilon}{\mathrm{max}}\ \ell(f_\theta(x+\delta ),z)\right] = 
\nabla_\theta\ \ell(f_\theta(x+\delta^\ast),z)
\end{equation}
with $\delta^\ast$ the solution of the maximisation problem\footnote{Such theorem is used rigorously  in reference \citep{2019arXiv190502175I}.}, but it is also used even if the convexity is not guaranteed, as experimentally it gives satisfactory numerical results.The second point has been discussed in \citep{2016arXiv161101236K}, and it results that  a mixture of unperturbed samples and adversarial samples is more effective. The introduction of adversarial samples acts like a regularisation term \citep{2018arXiv181000953F,2018arXiv181000363B}. 

So, with the tools developped in the previous section, we can set-up a new training algorithm (alg.~\ref{algo-1}) where after choosing the fraction of adversarial samples that populate each training mini-batch, one computes $\delta^\ast$ either by using FSGM one-step method  or by using the PGD iterative method. Then, one updates the gradients involved in the gradient descent step to find the loss minimum with respect to the model weights ($\theta$). This schema is rather simple and for instance the last step (12) is usually taken in charge by more sophisticated algorithms already used in classical training, i.e. SGD, Adam, and so on, with  for instance momentum memory and weight decay. Moreover, the attack generators listed  are only those presented in the previous section, and for other applications they can be modified on purpose.
\begin{table}
\caption{Results after an adversarial training (alg.~\ref{algo-1}) of an \textit{Inception} model using 100k images (non perturbed) among which a certain fraction ($f_a$) are perturbed using a FSGM method with $\varepsilon = 10^{-2}$.  For each $f_a$ value, are presented the values of the bias, the $\sigma_\mathrm{MAD}$ and $\eta$ as for \tab{\ref{tab-non-robust}}  for  the cases of  'non perturbed/perturbed' images obtained with the same attack that has been used for the training. For the special case where $f_a=0\%$, are reminded the values presented in \tab{\ref{tab-non-robust}} to ease the comparison and the bias value obtained with perturbed images are noted for completeness.}
\label{tab-robust}
\centering
\begin{tabular}{lr@{/}lr@{/}lr@{/}l} \hline \hline
fraction of adv. &  \multicolumn{2}{c}{bias ($\times 10^{-4}$)}   & 
 \multicolumn{2}{c}{$\sigma_\mathrm{mad}$ ($\times 10^{-3}$)} &  \multicolumn{2}{c}{$\eta$ (\%)} \\ \hline
$0\%$  & $-0.3$ & $-105$   &  $11$ & $66$  &   $1$ & $42$ \\ 
$5\%$  & $-20$ & $-40$      & $11$ & $9$    &  $1$&$4$ \\
$10\%$ & $40$&$-25$        & $11$&$8$      & $1$&$2$ \\
$20\%$ & $-6$&$23$         & $11$&$8$       & $1$&$1$\\
\hline
\end{tabular}
\end{table}
\subsection{Training with FSGM adversarial samples}
The results obtained after an adversarial training of an \textit{Inception} model using (step (7) of alg.~\ref{algo-1}) a FSGM attack generator  with $\varepsilon=10^{-2}$ are presented in \tab{\ref{tab-robust}}. We have varied the fraction ($f_a$)  of adversarial samples (i.e. perturbed images) among each batch training sets. It is quite striking that a rather small fraction of adversarial samples as $5\%$, already decreases the sensibility of the model to adversarial samples.  Increasing the fraction $f_a$ leads to better results with similar performances for non perturbed or perturbed images. We note however a worse bias compared to a classical training using non perturbed images. Investigating different optimiser codes and tweaking their own settings does not change drastically this behaviour. In passing, we notice that if the absolute values of the variables, notably the bias, are varying with the optimiser settings, the relative behaviour after training between non perturbed and perturbed images remains unchanged.  
\subsection{The PGD adversarial samples}
The results obtained in the previous section  using  the FSGM adversarial sample generator  seem encouraging,  but the \textit{Inception} model trained with $f_a=20\%$ or greater value of FSGM adversarial samples has no increase of robustness against the PGD attack using $\varepsilon = 10^{-2}$ and $n_{iter}=10$. This weakness is even still present when we use during the training a fraction of adversarial samples generated by the PGD iterative method. This problem has been  identified in reference \citep{2016arXiv161101236K} and an explanation has been elaborated  by the authors of \citep{2017arXiv170606083M}. In summary, they show that around a given sample $x$, the decision boundary should be modified in a complex manner in such a way that the perturbed samples $(x+\delta^\ast)$  are correctly classified. This is a sort of local underfitting problem. To do so, obtaining a stronger robustness against PGD iterative perturbations requires at least a model with a larger capacity. 

Considering the \textit{Inception} architecture (\tab{\ref{tab-inception-model}}), to get a larger capacity is not just a matter of increasing the number of model parameters (i.e. the length of the $\theta$ vector). For instance, by increasing the output size of the first fully connected layer, that is to say the 'fc0' layer and accordingly the input size of the 'fc1' layer (\tab{\ref{tab-inception-model}}), from 1096 to 2000 neurons, leads to a new CNN network with $48\ 722\ 852$ parameters compared to the original \textit{Inception} model with  $27\ 596\ 372$ parameters. However, this larger network has no better robustness against PGD attack. It is not neither a matter of increasing the depth of the convolution part. For instance, we have double the number of 'i3' inception cells without any success of increasing the robustness of the modified model. The only way we have found to manage a better robustness is by increasing the number of features entering the classifier, that is to say the input size of the 'fc0' layer. For instance,  we have increased the output size of the 'i4.s2\_2' layer from 128 to 256, then the 'fc0' input length has been increased from 22273 to 38657, leading a model with $45\ 720\ 660$ parameters. Pushing further the output size of the  'i4.s2\_2' layer 
does not change the conclusions. 
\begin{table}
\caption{Results of modified version (see text) of the \textit{Inception} baseline model (\tab{\ref{tab-inception-model}}) trained with 50\% of PDG adversarial samples ($\varepsilon = 10^{-2}$ and $n_{iter}=10$) and 50\% of non perturbed samples. To ease the comparison, we have reported from \tab{\ref{tab-non-robust}} the result obtained with the \textit{Iref} model. For each variables (bias, $\sigma_\mathrm{mad}$ and $\eta$) are given the results considering non perturbed samples, samples generated by FSGM method ($\varepsilon = 10^{-2}$) and samples generated by the PGD method used for training.
}
\centering
\begin{tabular}{lccc} \hline \hline
  Model & bias ($\times 10^{-4}$) & $\sigma_\mathrm{mad}$ ($\times 10^{-3}$) & $\eta$ (\%) \\ \hline
\textit{Iref} &
                 $0.3$/--/-- &  $11/66/82$    & $1/42/59$ \\ 
\textit{I(modified)} &
          $-21/-32/-32$ & $15/24/25$ & $2/6/6$ \\
\hline
\end{tabular}
\label{tab-PGD-training}
\end{table}

The results are shown  in \tab{\ref{tab-PGD-training}} with $f_a=50\%$. Clearly we have gained in robustness with comparable resistance against FSGM and PGD adversarial sample generations. But, in the same times we have degraded the results obtained for normal samples which is not satisfactory. Actually, using $f_a\leq10\%$ leads to results very similar the case of \textit{Iref} and increasing $f_a$ tends to increase the robustness against adversarial PGD attack and simultaneously worsen the results for normal samples. Notice that training with the PGD generator $n_{iter}$ parameter set to 30 do not change the results. The problem is probably due to the  architecture of the \textit{Inception} model. Although, testing over architecture is beyond the purpose of this article and is postponed to further studies. As suggested by reference \citep{2018arXiv181000144Y}, the inspection of the loss landscape in the input space rather than in the parameters space could be a diagnose of the susceptibility of a model to adversarial samples and a guideline to propose an alternative.
\section{Summary}
\label{sec:conclusion}
In this article we have conducted some numerical experiments with a specific convolutional neural network  (i.e. the \textit{Inception} model) specially trained to predict galaxy redshift from datacubes composed of SDSS multi-band  photometric imaging. We have shown that despite a good performance using "classical" training on normal images (i.e. non perturbed images), the model has a very poor generalisation power when considering very tiny, although special, perturbations of these images far below the background noise. This is the quintessence of adversarial images to find the best perturbation that fool the network prediction. These perturbed images reveal weaknesses that we think one should take care about, especially to study the robustness of the model. Notice that we would have drawn the same conclusions with an other CNN network, this is not specific to the \textit{Inception} model.

Following ML developments on this subject, we have presented  a simple algorithm for adversarial training (alg.~\ref{algo-1}) that injects during the training phase a fraction of adversarial images. This tends to strengthen the model against those adversarial images without degrading to much the results on normal images. This is especially efficient considering Fast Sign Gradient Method to generate adversarial images and we have shown that the present \textit{Inception} model can be well retrained. But, still the model does not master more aggressive adversarial generators such as the Projected Gradient Descent iterative methods. We have tried different modifications of the model to enlarge its capacity and found that increasing the number of features entering the classifier part of the network is the only way to get a better robustness to PGD generated samples, but it is at the price of the degradation of the redshift predictions with non perturbed images which is not satisfactory. This might be the sign that the \textit{Inception} model architecture is not well adapted to fight against such attack. However, we note that the adversarial training  alg.~\ref{algo-1} may also be improved to perform better inner maximisation, for instance using a network specially trained to generate adversarial samples \citep{Jang_2019_ICCV}. This is postponed to further studies.

As final remarks, we want to point that the sensitivity to adversarial samples is neither restricted to photometric redshift predictions and the other usages of neural networks in the different fields of image based analysis should be addressed, nor a specificity of convolutional neural networks as tree-based models like Decision Trees, Gradient Boosted DT, Random Forests may also be affected as described in reference \citep{2019arXiv190210660C}. 
\section*{Acknowledgements}
The author wants to warmly thank J. Pasquet-Itam and E. Bertin for the fruitful discussions and the access to the SDSS data set, as well as the details of the \textit{Inception} model design they have provided to me.

\bibliographystyle{mnras}
\bibliography{papier}

%

\appendix
\section{\textit{Inception} and CNN architectures}
\label{app-inception}
In this appendix, we give a more advanced description of the different layers of the two models used in this article:  the \textit{Inception} CNN used by \citep{2019A&A...621A..26P} (\tab{\ref{tab-inception-model}}), and a simpler CNN architecture (\tab{\ref{tab-cnn-model}}).

\onecolumn
\begin{table}
\caption{\textit{Inception} scheme with a total number of trainable parameters $27\ 596\ 372$. Kernel sizes of convolution layers are detailed as well as padding if not 1x1. 'pool0' are average pooling layers with possibly associated padding layers. Concatenation of the different layers are mentioned in inception cells ('i0',...,'i4') as well as the first fully connected layer ('fc0') with 'ebv'  information added. All the 's1\_0',  's1\_2',  's1\_1',  's2\_2' layers share the same convolutional structure. The tensor dimensions follow the pyTorch convention $Channel \times Height \times Width$ for non-fully connected layers for which flatten tensors are used.}
\label{tab-inception-model}
\begin{tabular}{{lllrr}}
\hline\hline
          name &                     input\_shape &         output\_shape &  nb\_params & Comments\\
\hline
         conv0 &               (5, 64, 64) &   (64, 64, 64) &       8128 & kernel=(5, 5), padding=(2, 2)\\
         pool0 &              (64, 64, 64) &   (64, 32, 32) &          0 & average\\
\hline
       i0.s1\_0 &              (64, 32, 32) &   (48, 32, 32) &       3168 & kernel=(1, 1)\\
 \phantom{xx}      i0.s2\_0 &              (48, 32, 32) &   (64, 32, 32) &      27776 & kernel=(3, 3)\\
      i0.s1\_2 &              (64, 32, 32) &   (48, 32, 32) &       3168 & kernel=(1, 1)\\
  \phantom{xx}     i0.pool0 &              (48, 33, 33) &   (48, 32, 32) &          0 &\\
       i0.s1\_1 &              (64, 32, 32) &   (48, 32, 32) &       3168 &  kernel=(1, 1)\\
   \phantom{xx}     i0.s2\_1 &              (48, 32, 32) &   (64, 32, 32) &      76928 & kernel=(5, 5), padding=(2, 2)\\
       i0.s2\_2 &              (64, 32, 32) &   (64, 32, 32) &       4224 & kernel=(1, 1)\\
            i0 &              (64, 32, 32) &  (240, 32, 32) &     118432 & concat i0.(s2\_2, s2\_1, s2\_0, pool0)\\
\hline
       i1.s1\_0 &             (240, 32, 32) &   (64, 32, 32) &      15488 & \\
  \phantom{xx}       i1.s2\_0 &              (64, 32, 32) &   (92, 32, 32) &      53176 &\\
       
       i1.s1\_2 &             (240, 32, 32) &   (64, 32, 32) &      15488 & \\
 \phantom{xx}       i1.pool0 &              (64, 33, 33) &   (64, 32, 32) &          0 &\\
 
       i1.s1\_1 &             (240, 32, 32) &   (64, 32, 32) &      15488 & \\
 \phantom{xx}        i1.s2\_1 &              (64, 32, 32) &   (92, 32, 32) &     147384 & \\
       i1.s2\_2 &             (240, 32, 32) &   (92, 32, 32) &      22264 & \\
            i1 &             (240, 32, 32) &  (340, 32, 32) &     269288 & concat i1.(s2\_2, s2\_1, s2\_0, pool0)\\
         pool0 &             (340, 32, 32) &  (340, 16, 16) &          0 &\\
\hline
       i2.s1\_0 &             (340, 16, 16) &   (92, 16, 16) &      31464 &\\
 \phantom{xx}        i2.s2\_0 &              (92, 16, 16) &  (128, 16, 16) &     106240 &\\

       i2.s1\_2 &             (340, 16, 16) &   (92, 16, 16) &      31464 &\\
 \phantom{xx}       i2.pool0 &              (92, 17, 17) &   (92, 16, 16) &          0 &\\
 
       i2.s1\_1 &             (340, 16, 16) &   (92, 16, 16) &      31464 &\\
 \phantom{xx}        i2.s2\_1 &              (92, 16, 16) &  (128, 16, 16) &     294656 &\\
       i2.s2\_2 &             (340, 16, 16) &  (128, 16, 16) &      43776 &\\
            i2 &             (340, 16, 16) &  (476, 16, 16) &     539064 & concat i2.(s2\_2, s2\_1, s2\_0, pool0)\\
\hline
       i3.s1\_0 &             (476, 16, 16) &   (92, 16, 16) &      43976 &\\
 \phantom{xx}        i3.s2\_0 &              (92, 16, 16) &  (128, 16, 16) &     106240 &\\
       i3.s1\_2 &             (476, 16, 16) &   (92, 16, 16) &      43976 &\\
 \phantom{xx}       i3.pool0 &              (92, 17, 17) &   (92, 16, 16) &          0 &\\

       i3.s1\_1 &             (476, 16, 16) &   (92, 16, 16) &      43976 &\\
 \phantom{xx}        i3.s2\_1 &              (92, 16, 16) &  (128, 16, 16) &     294656 &\\
       i3.s2\_2 &             (476, 16, 16) &  (128, 16, 16) &      61184 &\\
            i3 &             (476, 16, 16) &  (476, 16, 16) &     594008 & concat i3.(s2\_2, s2\_1, s2\_0, pool0)\\
         pool0 &             (476, 16, 16) &    (476, 8, 8) &          0 &\\
\hline 
       i4.s1\_0 &               (476, 8, 8) &     (92, 8, 8) &      43976 &\\
 \phantom{xx}        i4.s2\_0 &                (92, 8, 8) &    (128, 8, 8) &     106240 &\\
       i4.s1\_2 &               (476, 8, 8) &     (92, 8, 8) &      43976 &\\
 \phantom{xx}       i4.pool0 &                (92, 9, 9) &     (92, 8, 8) &          0 &\\
       i4.s2\_2 &               (476, 8, 8) &    (128, 8, 8) &      61184 &\\
            i4 &               (476, 8, 8) &    (348, 8, 8) &     255376 & concat i4.(s2\_2, s2\_0, pool0)\\
\hline
           fc0 &                   (22273) &         (1096) &   24412304 & concat (i0,i1,i2,i3,i4, 'ebv')\\
           fc1 &                    (1096) &         (1096) &    1202312 &\\
           fc2 &                    (1096) &          (180) &     197460 &\\
\hline
\end{tabular}
\end{table}
\begin{table}
\caption{\textit{CNN} scheme with a total number of trainable parameters $13\ 461\ 008$. Kernel sizes of convolution layers are detailed as well as padding if not 1x1. 'pool0' are average pooling layers with possibly associated padding layers. The last convolutional-polling associated layers ('pool2')'  output is concatenated with the 'ebv'  information added to feed the first fully connected layer. The tensor conventions follow those describes in \tab{\ref{tab-inception-model}}.}
\label{tab-cnn-model}
\begin{tabular}{lllrr}
\hline\hline
        name &                     input\_shape &         output\_shape &  nb\_params & Comments\\
\hline
       conv0 &               (5, 64, 64) &   (64, 64, 64) &       8064 & kernel=(5, 5), padding=(2, 2)\\
       \phantom{xx} pool0 &              (64, 64, 64) &   (64, 32, 32) &          0 & average\\
       conv1 &              (64, 32, 32) &   (92, 34, 34) &      53084 & kernel=(3, 3), padding=(2, 2)\\
       \phantom{xx} pool1 &              (92, 34, 34) &   (92, 17, 17) &          0 & average\\
       conv2 &              (92, 17, 17) &  (128, 19, 19) &     106112 & kernel=(3, 3), padding=(2, 2)\\
       \phantom{xx} pool2 &             (128, 19, 19) &  (128, 10, 10) &          0 & average \\
         fc0 &                   (12801) &         (1024) &   13109248 & concat (pool2, 'ebv')\\
         fc1 &                    (1024) &          (180) &     184500 \\
\hline
\end{tabular}
\end{table}
%

\bsp	
\label{lastpage}
\end{document}